\documentclass[11pt]{article}

\usepackage[a4paper,margin=2.15cm]{geometry}
\usepackage[T1]{fontenc}
\usepackage[utf8]{inputenc}
\usepackage{lmodern}
\usepackage{xcolor}
\usepackage{comment}
\usepackage{microtype}
\usepackage{amsmath,amssymb,bm}
\usepackage{graphicx}
\usepackage{booktabs}
\usepackage[numbers,sort&compress]{natbib}
\usepackage[colorlinks=true,linkcolor=blue,citecolor=blue,urlcolor=blue]{hyperref}
\usepackage{authblk}
\graphicspath{{figures/}}
\usepackage[left]{lineno}

\title{\textbf{Topologically protected chiral sensing using Synthetic Chiral Light}
}
\author[1,2]{Gefei Li\thanks{These authors contributed equally to this work.}}
\author[3]{Justas Terentjevas\textsuperscript{*}}
\author[4]{Yong Zhang}
\author[5]{Patricia Vindel-Zandbergen}
\author[6]{David Ayuso}
\author[3]{Serguei Patchkovskii}
\author[7]{Tran Tien Dat}
\author[4]{Junpeng Lu}
\author[4]{Qihua Liu}
\author[1,2]{Yuanjie Pan}
\author[1,2]{Li Liu}
\author[1,2]{Hao Teng}
\author[1,2]{Zhiyi Wei}
\author[3,8,9]{Misha Yu. Ivanov}
\author[3,9,10]{Olga Smirnova\thanks{smirnova@mbi-berlin.de}}
\author[1,2]{Pengju Zhang\thanks{pengju.zhang@iphy.ac.cn}}
\date{}

\affil[1]{Institute of Physics, Chinese Academy of Sciences, Beijing 100190, China}
\affil[2]{School of Physical Sciences, University of Chinese Academy of Sciences, Beijing 100190, China}
\affil[3]{Max Born Institute for Nonlinear Optics and Short Pulse Spectroscopy, 12489 Berlin}
\affil[4]{School of Electronic Science \& Engineering, Southeast University, Nanjing, China}
\affil[5]{Department of Chemistry, New York University, New York 10003, New York, USA}
\affil[6]{Department of Chemistry, Molecular Sciences Research Hub, Imperial College London, W12 0BZ London, UK}
\affil[7]{Department of Physics, University of Hong Kong, Hong Kong SAR 999077, China}
\affil[8]{Department of Physics, Humboldt Universität zu Berlin, Newtonstrasse 15, Berlin 12489, Germany}
\affil[9]{Solid State Institute, Technion, Israeli Institute of Technology, Haifa, Israel}
\affil[10]{Technische Universität Berlin, 10623 Berlin, Germany}

\begin{document}


\maketitle

\begin{abstract}

Chirality underlies molecular function in living matter, yet its optical detection remains challenging because conventional chiroptical spectroscopies rely on weak corrections to the dominant electric-dipole light-matter interaction, making desired optical signals weak and fragile. Topology offers a route to robustness, enabling observables whose defining properties survive disorder and imperfections. However, experimental realization of a  practical topological observable for chiral spectroscopy has remained elusive.

Here we realize chiral topological light and demonstrate such an observable. 
A tightly focused, phase-locked, counter-rotating two-colour field encodes chirality in the three-dimensional  electric-field   trajectory while its dominant  topological charge resides in the longitudinal electric-field component where it remains hidden from direct far-field detection. 
An isotropic chiral medium acts as a topological transducer, converting the latent topology of the driving field into a propagating nonlinear response whose topological charge becomes directly observable in the far field while remaining strongly suppressed in achiral media.   Using randomly oriented chiral single-crystal powders, we experimentally detect the enantio-sensitive response through  the topological charge of the emitted field.
Directly accessible in the far field and robust against experimental imperfections, this observable allows us to track and control chiral signal on attosecond timescales through the relative phase of the driving two-colour fields. 

Our results establish topology as a practical resource for ultrafast chiral optical spectroscopy.
\end{abstract}

Chirality gives rise to enantiomers — left- and right-handed geometric forms of matter whose imbalance governs processes ranging from molecular recognition and pharmaceutical efficacy \cite{Barron2004,Nguyen2006,Bitchagno2022} to metabolism and biological function \cite{liu2023detection}. Yet while nature exploits chirality with remarkable reliability, 
conventional chiroptical spectroscopies rely on weak magnetic-dipole and electric-quadrupole light-matter interactions \cite{Berova2012,oppermann2024capturing,changenet2023recent}, yielding signals that are often orders of magnitude smaller than their electric-dipole counterparts \cite{Ayuso2022PCCP,smirnova2025new} and therefore particularly susceptible to noise, disorder, and experimental imperfections. The challenge becomes particularly acute in ultrafast time-resolved optical detection \cite{oppermann2019broad}, discriminating small chiral molecules \cite{liu2023detection}, powders and condensed-phase systems where disorder,  depolarization, and optical artifacts further complicate reliable chiral detection \cite{Harada2013}. These limitations have motivated intense efforts to enhance chiral light–matter interactions using structured optical fields \cite{forbes2018optical, Ayuso2019, forbes2021orbital,begin2023nonlinear,rouxel2022hard,hrast2026bottom, Mayer2024}. 

Topology offers a route from geometry to robustness, protecting observables against perturbations while preserving their sensitivity to underlying geometric structure.  Optical vortex beams carrying orbital angular momentum (OAM) \cite{Allen1992,Shen2019} can influence chiral observables in a way sensitive to the sign and magnitude of a beam's topological charge \cite{forbes2018optical,forbes2021orbital,begin2023nonlinear,rouxel2022hard,hrast2026bottom}. 
However, in these approaches topology primarily acts as a property of the driving field rather than as a mechanism that protects the measured chiral response. Furthermore, OAM-based schemes generally require vortex beams  and often rely on field gradients or higher-order light–matter couplings \cite{forbes2018optical,forbes2021orbital,begin2023nonlinear,rouxel2022hard,hrast2026bottom}. As a result, the experimental realization of robust topological observables for chiral detection has remained elusive.
\begin{figure*}[t]
\centering
\includegraphics[width=0.90\textwidth]{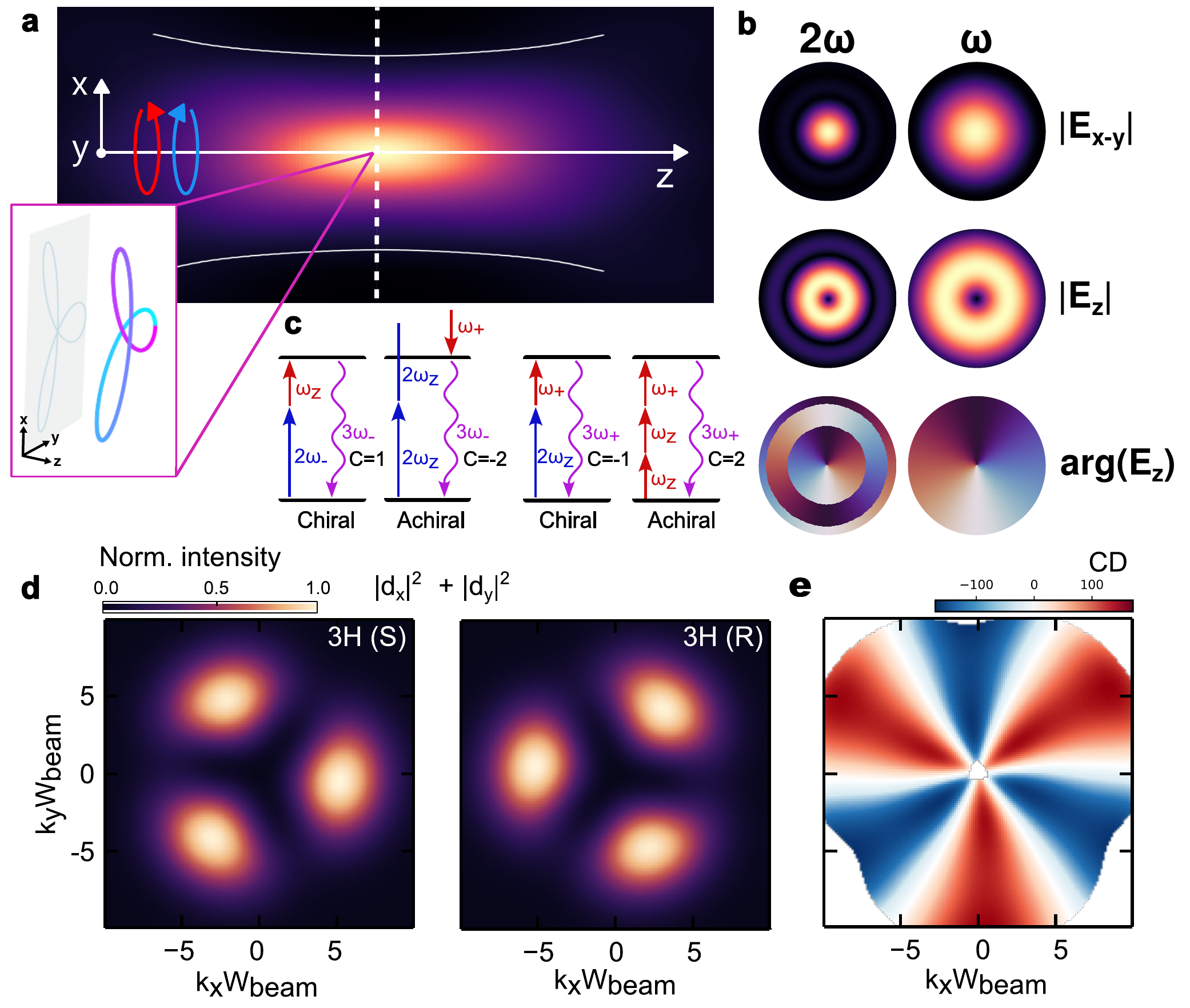}
\caption{\textbf{Chiral topological light via tight focusing.} \textbf{a--b}, Counter-rotating two-colour circularly polarized fields form a trefoil trajectory whose orientation is set by the relative phase of the \(\omega\) and \(2\omega\) components. Tight focusing generates a longitudinal field leading to three-dimensional Lissajous figure with the handedness and azimuthal winding defined by the relative spin of the \(\omega\) and \(2\omega\) beams. Field components are shown in the focal plane. \textbf{c}, Multiphoton  chiral and achiral pathways leading to $3\omega_{\pm}$ emission, where subscript $\pm$ denotes the spin angular momentum of the emitted field. The integer \(C\) denotes the angular winding carried by the corresponding pathway. Simulated far-field third-harmonic intensity  \textbf{d}, and chiral dichroism \textbf{e}, for randomly oriented carvone molecules ($\lambda_{+}=800~\mathrm{nm}$ and $\lambda_{-}=400~\mathrm{nm}$, $I_{+}=I_{-}=10^{12}~\mathrm{W/cm^2}$ and $\mathrm{FWHM}_{+}=\mathrm{FWHM}_{-}=12.5~\mathrm{fs}$). }
\label{fig:field}
\end{figure*}
A different route was recently proposed through synthetic chiral light, a class of locally chiral ultrafast structured optical fields that encodes chirality in the three-dimensional electric field trajectory and employs the strongest, electric-dipole way of light–matter coupling mapping molecular handedness  into nonlinear optical signals \cite{Ayuso2019,kohnke2025multiphoton}. 
Correlating local light chirality  across both time and azimuthal space one can enable chiral topological light \cite{Mayer2024}. Its handedness is defined locally by the three-dimensional temporal orbit of the electric field, while the collective arrangement of these orbits around the beam axis generates a chiral vortex whose winding number is fixed by the topology of the contributing light modes \cite{Mayer2024}. While a local distortion may influence the shape of the orbit, it cannot continuously remove the winding preserving topological properties encoded in the angular shape. For a nonlinear measurement, this means that the material response is encoded in the angular phase of the signal. This is particularly beneficial for powders, heterogeneous solids and liquids, where total intensity measurements may be unreliable. 

Here we experimentally realize synthetic chiral light whose hidden topology is revealed through its interaction with chiral matter, establishing the first experimental realization of chiral topological light and enabling robust enantio-sensitive optical detection.  We show that topological optical signatures can arise as an intrinsic property of chiral media rather than being inherited from optical vortex beams in the driving field. 
\begin{figure*}[t]
\centering
\includegraphics[width=0.98\textwidth]{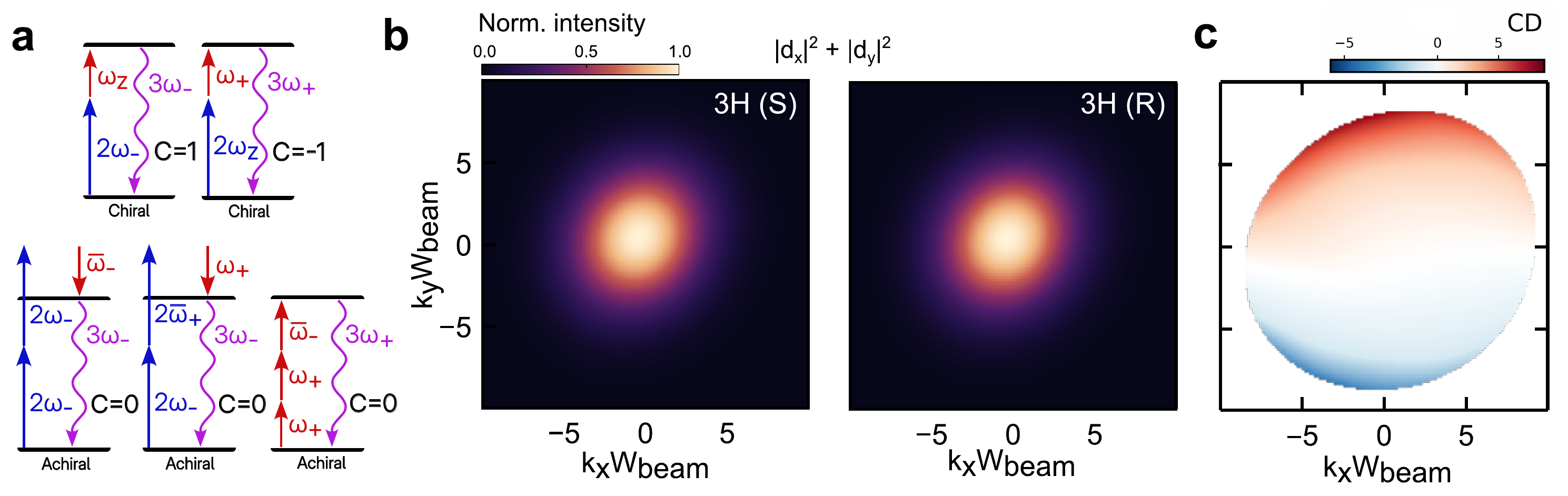}
\caption{\textbf{Chiral topological light driven by elliptical two-colour fields.} \textbf{a}, Multiphoton chiral and achiral pathways leading to $3\omega_{\pm}$ emission for elliptically polarized driving field. $\overline{\omega}_{\pm}$ denotes weak "counter-rotating" components responsible for the emergence of elliptical polarization. \textbf{b}, Far-field intensity of the third harmonic emitted from randomly oriented gas phase carvone molecules ($\epsilon_{+}=\epsilon_{-}=0.95$,  $\lambda_{+}=800~\mathrm{nm}$ and $\lambda_{-}=400~\mathrm{nm}$, $I_{+}=I_{-}=10^{12}~\mathrm{W/cm^2}$ and $\mathrm{FWHM}_{+}=\mathrm{FWHM}_{-}=12.5~\mathrm{fs}$). \textbf{c}, Corresponding chiral dichroism: $2\frac{I_{R}(\rho,\theta)-I_{S}(\rho,\theta)}{I_{R}(\rho,\theta)+I_{S}(\rho,\theta)}$. The azimuthal intensity pattern is consistent with a topological charge of $C=\pm1$ and is rotated by $\pi$ between enantiomers. Because the chiral response resides in a distinct angular harmonic set by the topological charge, it can be extracted in a background-free manner through Fourier analysis with respect to the two-colour delay. 
}
\label{fig:calculation}
\end{figure*}
The physical origin of the chiral-topological response is most transparent for a pair of tightly focused two-colour counter-rotating circularly polarized fields $\omega-2\omega$ (Figure~\ref{fig:field}a). Tight focusing plays a dual role. First, it generates a longitudinal electric-field component. Together with the transverse components of the fundamental and second-harmonic fields, this longitudinal field provides the third non-coplanar polarization direction required for synthetic chiral light \cite{Ayuso2019}, enabling the electric field to trace a locally chiral three-dimensional trajectory in time (Figure~\ref{fig:field}b). Second, focusing converts part of the spin angular momentum of circularly polarized light into orbital angular momentum \cite{Bliokh2015}, endowing the longitudinal component with an azimuthal phase winding ($e^{i\sigma\theta}$), where ($\sigma=\pm1$) denotes the light spin\footnote{Focusing also generates a much weaker transverse component carrying orbital angular momentum, whose amplitude scales as $\sin^2{\alpha/2}$, where $\alpha$ is the focusing half-angle \cite{Bliokh2015}. This contribution is negligible in our experimental geometry.}. The longitudinal field therefore carries a topological charge ($C=\sigma$), despite the absence of optical vortex beams in the incident field.

This topology remains hidden because the longitudinal field does not directly contribute to the observable far-field emission. However, its presence becomes apparent through nonlinear light–matter interactions (Figure~\ref{fig:field}c). The lowest-order chiral pathway corresponds to sum-frequency generation \cite{giordmaine1965,belkin,fischer} resulting in emission of $3\omega$ field and involves a single longitudinal photon.  It therefore inherits one unit of topological charge from the driving field. By contrast, the lowest-order achiral pathway contains two longitudinal photons and carries two units of topological charge.  The weakness of the longitudinal field suppresses achiral pathways relative to chiral ones, allowing chirality to act as the efficient transducer of hidden optical topology into an observable far-field response.

The chiral (C=±1) and achiral (C=±2) pathways emit third-harmonic light $3\omega_{\pm}$ with  spin angular momentum $\pm1$ indicated as subscript (Fig.~\ref{fig:field}c). Their interference produces an enantio-sensitive azimuthal intensity modulation carrying topological charge $C=\pm 3$.
To illustrate the mechanism in an isotropic chiral medium, we numerically solve the time-dependent Schrödinger equation (TDSE) for randomly oriented carvone molecules, modelling the tightly focused driving fields using the Richards--Wolf formalism (see Methods). Figure~\ref{fig:field}d,e shows the resulting far-field intensity of the third harmonic for the S and R enantiomers. The emission exhibits a pronounced 3-fold angular structure indicating the underlying topological charge $C=\pm3$. The $\pi$-phase shift of the chiral response between opposite enantiomers is converted by the topological charge into a rigid rotation of the far-field pattern by $\Delta\theta=\pi/C=\pi/3$ \cite{Mayer2024}. 
The angular phase therefore directly encodes molecular handedness while remaining tied to the underlying topology of the nonlinear response. The observed topological charge is fully consistent with the general chiral-topological framework of Ref.~\cite{Mayer2024}, demonstrating that non-trivial topology can emerge in the enantio-sensitive response even in the absence of orbital angular momentum in the driving beams.

Chiral topological light exhibits distinct topological structures for circular and elliptical polarizations \cite{Mayer2024};  we now focus on the more general and experimentally relevant case of counter-rotating elliptically polarized fields with ellipticity $\epsilon=0.95$, which is typical for our experiments. Ideal circular polarization is fundamentally fragile in practice; cumulative Fresnel amplitude and phase shifts between the $s$ and $p$ components across standard optical elements—such as mirrors, windows, and waveplates—inevitably degrade the beam into a slightly elliptical state. Unlike the circularly polarized limit, ellipticity opens additional achiral pathways that do not require longitudinal components and therefore dominate the third-harmonic emission (Fig.\ref{fig:calculation}a). Nevertheless, the enantio-sensitive response remains tied to the longitudinal-field phase winding, allowing the hidden topology generated via focusing to be transferred selectively to the chiral component of the emitted field.
The topological origin of the resulting response becomes apparent at the level of multiphoton pathways (Fig.~\ref{fig:calculation}a). 
Chiral-sensitive processes, such as sum-frequency generation \cite{giordmaine1965, belkin,fischer}, necessarily involve these longitudinal channels and therefore transfer their otherwise hidden topology to the emitted transverse ($3\omega$) field. In contrast, the achiral nonlinear processes, including conventional third-harmonic generation and four-wave mixing, do not require longitudinal fields and consequently the dominant achiral emission does not carry the topological charge in this case.

To illustrate the universal topological response of chiral matter, we consider randomly oriented gas-phase carvone molecules as a representative example. Figure~\ref{fig:calculation}b shows the TDSE simulations for the far-field  third-harmonic emission for the randomly oriented (S)- and (R)-enantiomers driven by chiral topological light with  elliptically polarized fields. While the total signal is dominated by achiral emission, the enantio-sensitive response preserves a distinct topological charge $C=\pm1$ (Fig.~\ref{fig:calculation}c). This  topological structure becomes apparent in the chiral dichroism signal (Fig.~\ref{fig:calculation}c) and can subsequently be extracted from the total emission in a background-free fashion. 
The chiral dichroism presents a significant fraction (8\%) of the average signal - hallmark of electric-dipole based chiral detection \cite{Patterson2013,Bowering2001,Powis2000,nahon2006,baumert2015, ayuso2022ultrafast,smirnova2025new}. Importantly, as we demonstrate below, this topological response is robust against realistic experimental imperfections, including Gaussian beam distortions and beam misalignment, which are explicitly incorporated into the simulations using experimentally measured parameters (see Methods).

The two-colour phase delay controls the handedness of chiral topological light and provides direct access to the molecule-specific phase difference between chiral and achiral nonlinear pathways (see Methods). Fourier analysis of the far-field intensity with respect to this delay isolates the first harmonics ($m=\pm1$), yielding a background-free chiral observable. These components carry topological charge $C=\pm1$, corresponding to the third-harmonic emission. Their azimuthal phase encodes the molecular quantum phase and differs by $\pi$ between opposite enantiomers. Below we reconstruct this phase from the experiment.

To test whether the angular topological readout survives disorder and morphology-dependent optical artifacts, we used chiral hybrid perovskite powder films, $(R/S)$-3BrMBA$_2$PbI$_4$, prepared from enantiopure $(R/S)$-3BrMBA spacers. In these materials, molecular chirality is transferred from the organic spacers to the inorganic framework, where excitonic resonances, helical lattice distortions, and spin–orbit–coupled electronic states collectively govern the chiroptical response \cite{Jana2020,li2024large}. Powder samples provide a particularly demanding setting for chiral spectroscopy, as random crystallite orientations and anisotropy-induced artifacts tend to obscure conventional enantio-sensitive signals \cite{Harada2013}. They therefore constitute an ideal benchmark for testing whether a topological observable can preserve chiral sensitivity under realistic experimental conditions.

\begin{figure*}
\centering
\includegraphics[width=0.92\textwidth]{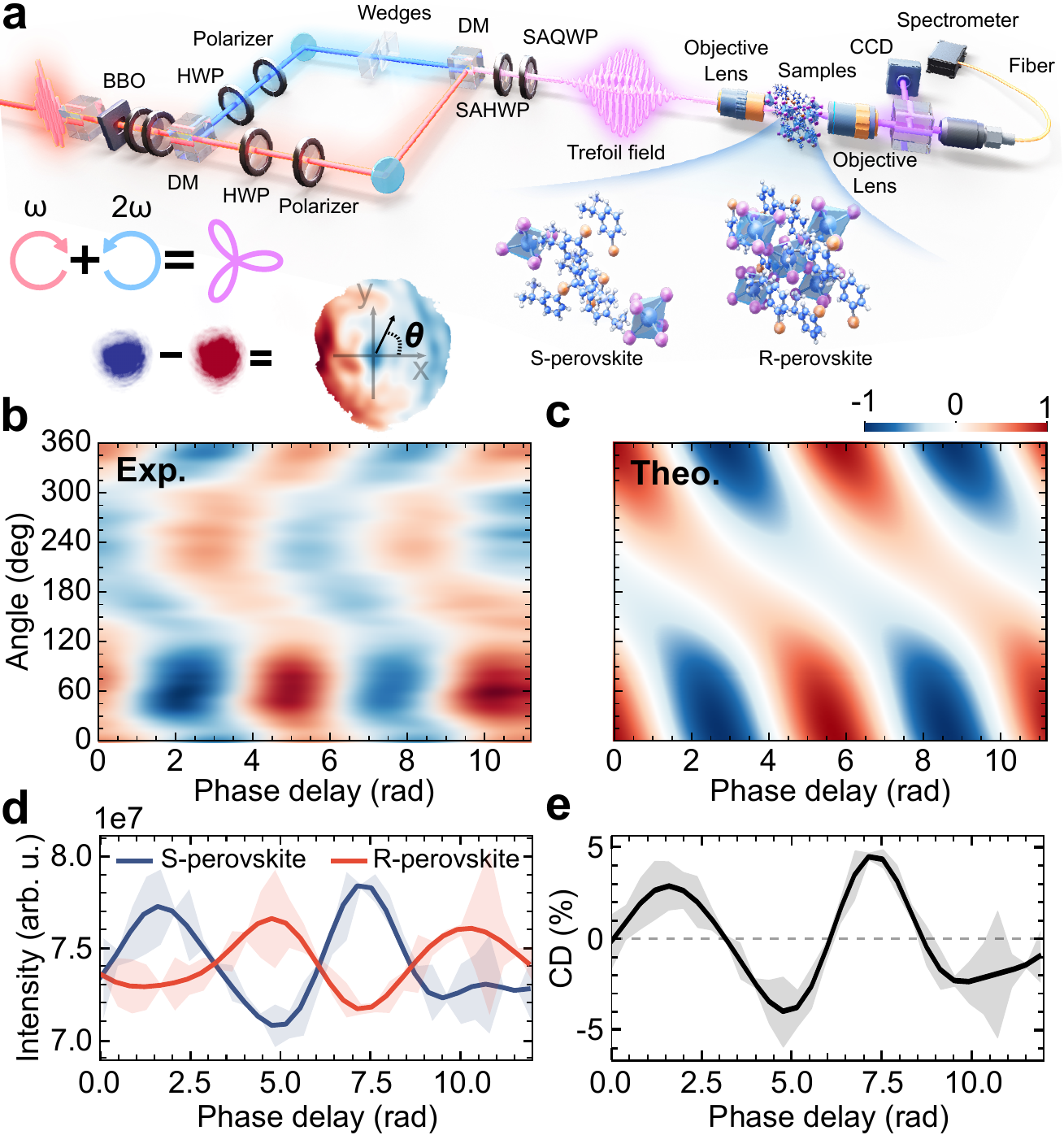}
\caption{\textbf{Enantiomer discrimination in chiral perovskite powders driven by a phase-locked trefoil field.}
\textbf{a}, Experimental scheme for generating synthetic chiral light by combining tightly focused counter-rotating circularly polarized fundamental and second-harmonic fields. The relative two-colour phase is controlled by wedges. The emitted third-harmonic signal is recorded simultaneously by Fourier-plane imaging and fiber-coupled spectral detection. HWP: half-wave plate; DM: dichroic mirror; QWP: quarter-wave plate; SAHWP: super-achromatic half-wave plate; SAQWP: super-achromatic quarter-wave plate.
\textbf{b}, Experimental angle--phase map obtained from the normalized difference between the Fourier-plane third-harmonic images of the two mirror-related powders. The schematic above this figure specifies the azimuthal angle and illustrates how the angular texture is extracted by presenting the experimental value of chiral dichroism resolved on azimuthal angle and radial coordinate. 
\textbf{c}, Calculated azimuthal angle--two-colour phase map under excitation by synthetic chiral light.  
\textbf{d}, Phase-dependent third-harmonic spectral intensities from $S$- and $R$-perovskite powder films.
\textbf{e}, Corresponding spectral chiral-dichroism trace.}
\label{fig:perovskite}
\end{figure*}

The measurement scheme is shown in Fig.~\ref{fig:perovskite}a. A focused, phase-locked $\omega$-$2\omega$ trefoil field, generated from a $2~\mu\mathrm{m}$ fundamental laser, drove third-harmonic emission from the $R$- and $S$-perovskite films. The peak electric-field amplitudes of the fundamental and second-harmonic components were set to approximately $8~\mathrm{MV~cm^{-1}}$. The relative two-colour phase, $\phi$, was scanned by translating the wedge pair, while the sample position, focusing condition and collection geometry were kept fixed. For each value of $\phi$, we recorded the third-harmonic emission in the Fourier plane, $I_{\eta}(\rho,\theta,\phi)$, where $\eta$ denotes the material handedness. This geometry preserves the azimuthal structure of the nonlinear emission, so a chiral response appears not only as a change in yield but also as a phase-locked angular texture. We therefore treated the angular phase of the emission as the primary enantio-sensitive observable. The following analysis compares the two enantiomers at the same optical phase, so that static optical backgrounds common to both measurements are suppressed before the angular texture is extracted.
 
To isolate the enantio-sensitive contribution, we compare the Fourier-plane images of the R and S powders acquired at the same optical phase. Their normalized difference (inset in Fig.~\ref{fig:perovskite}b) suppresses the common emission background, which likely arises from sample-to-sample variations in film thickness and reveals the $C=1$ component that changes sign with molecular handedness. The close correspondence between the inset of Fig.~\ref{fig:perovskite}b and the calculated dichroism in Fig.~\ref{fig:calculation}c confirms that the experimental difference map captures the $C=1$ angular winding predicted for the chiral nonlinear response. The red-blue texture follows a single angular period around the Fourier plane and remains phase-locked to the two-colour delay, as expected when the chiral component inherits one unit of azimuthal winding from the chiral topological light. Thus, the inset provides the first experimental signature of the topology-carrying response identified in the calculation. 

Next we integrate the images over radial dimension (see Methods) and present  an angle--phase map of the difference in angular resolved signals between the two enantiomers (Fig.~\ref{fig:perovskite}b), allowing the evolution of the chiral response to be tracked as the two-colour phase controlling the handedness of synthetic chiral light is scanned.

The resulting map provides direct evidence for enantiomer discrimination. The red and blue lobes drift continuously with the two-colour phase, demonstrating that the angular structure of the nonlinear emission is locked to the handedness of the synthetic chiral light. This phase-dependent winding is reproduced by the TDSE simulations shown in Fig.~\ref{fig:perovskite}c, supporting its origin in the interference of chiral and achiral nonlinear pathways (see Methods). 

Interestingly, both theory and experiment also exhibit a weaker enantio-sensitive modulation of the integrated harmonic yield: integration of signals in Figs.~\ref{fig:perovskite}b,c over azimuthal angle results in non-vanishing signal as a function of phase delay (Fig.~\ref{fig:perovskite}e). This residual response arises from the finite and generally non-uniform angular sampling of the topological far-field pattern. Experimental realities—such as fiber coupling, finite numerical apertures, and spectrometer slits—act as an inherent spatial filter, breaking the integration symmetry across the Fourier plane. Crucially, while chiral information is intrinsically encoded within the emission topology itself, this detection-induced projection is essential to map the underlying angular contrast into a measurable net intensity difference. This mechanism provides a highly accessible chiral observable for finite-aperture measurements. Consequently, scanning the two-colour relative phase drives a periodic, phase-shifted intensity modulation in the enantiomeric powders, yielding the distinct sign-reversing contrast.

\begin{figure}
\centering
\includegraphics[width=0.94\textwidth]{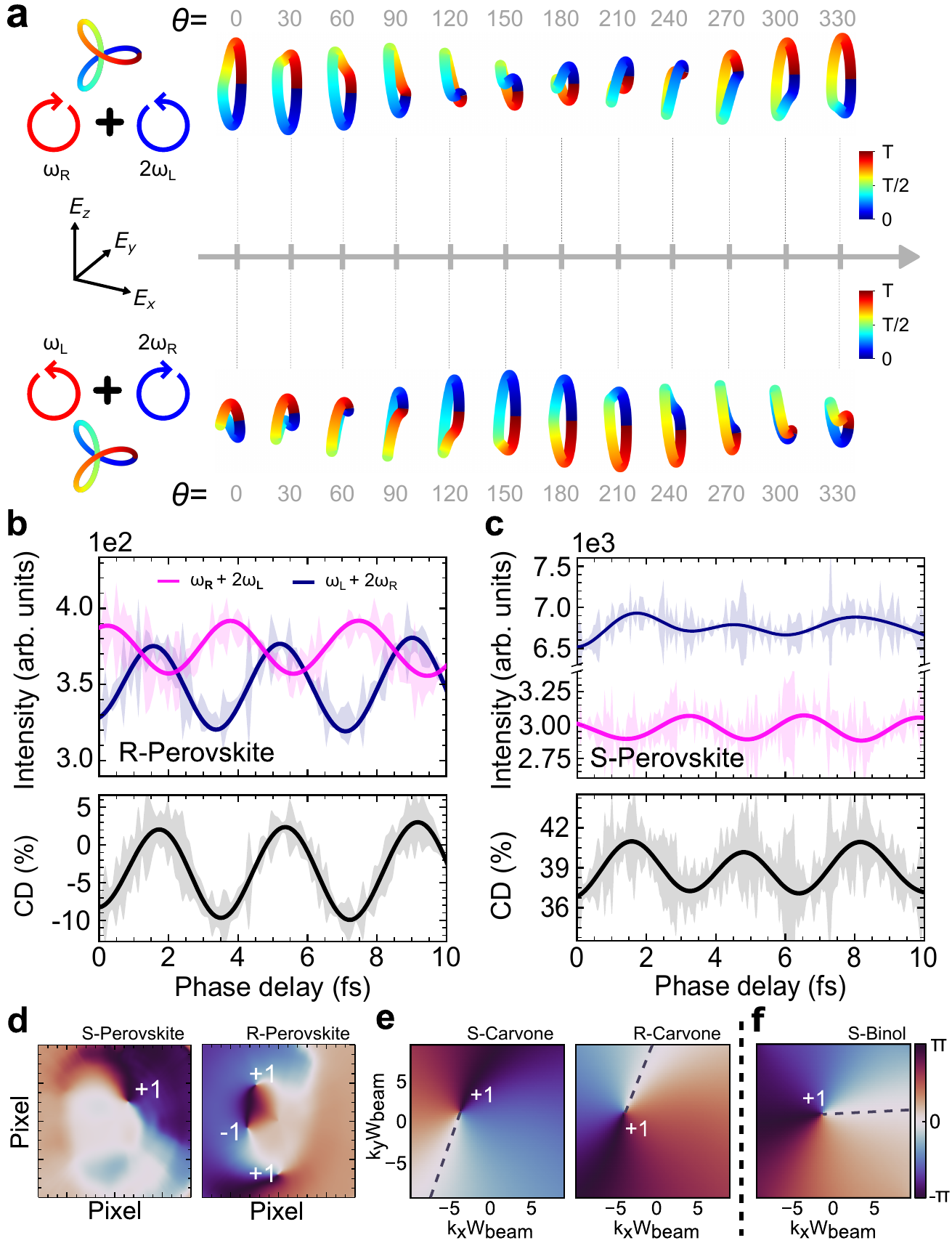}
\caption{\textbf{Isolating topological charge in third-harmonic emission.}
\textbf{a}, Chiral Lissajous figure of synthetic chiral light as a function of two-colour delay.
\textbf{b,c}, Spectrally selected third-harmonic intensity from R- and S-perovskite powders for the synthetic chiral light with the two opposite handedness configurations, \(\omega_R+2\omega_L\) and \(\omega_L+2\omega_R\), together with the corresponding normalized spectral contrast. \textbf{d}, Experimental far-field angular phase maps of the third-harmonic emission from the S- and R-perovskite powders obtained by Fourier transforming the far-field images with respect to two-colour phase. The angular phase maps corresponding to Fourier component $m=+1$ show enantiomer-dependent vortex-like structures in the emitted field with topological charge $C=+1$.
\textbf{e, f}, Corresponding calculated far-field angular phase maps for the two enantiomers of carvone and S-binol. Dashed line follows zero angular phase in all images to highlight the molecular-specific angular phase rotation.}
\label{fig:spectral_control}
\end{figure}

Having established a robust topological observable for chiral discrimination, we next examine the spectrally resolved third-harmonic yield to assess the strength and stability of the underlying nonlinear pathways. We use two distinct  configurations of synthetic chiral light, ($\omega_R+2\omega_L$) and ($\omega_L+2\omega_R$), corresponding to opposite handed optical trajectories and therefore driving different combinations of chiral interference channels, as illustrated in Fig.~\ref{fig:spectral_control}a. After spectral selection of the third-harmonic emission, both perovskite enantiomers retain a clear periodic dependence on the two-colour phase, confirming that the observed modulation originates from coherent nonlinear emission rather than from imaging artifacts.

At the same time, exchanging the handedness of the driving fields modifies the overall signal level and the achiral background contribution. This behaviour reflects the practical challenge of reproducing identical field structures when reversing the optical trajectory, as small differences in the relative amplitudes of the $\omega$ and 2$\omega$ components alter the balance between the contributing nonlinear pathways.

The mean intensities and contrast offsets differed substantially between the two enantiomeric films (Figs.~\ref{fig:spectral_control}b,c). We treat this difference as an important limitation of intensity-based chiral sensing rather than as the main observable. Small changes in film morphology, local thickness, field balance or collection efficiency can rescale the handedness-even background and therefore shift the apparent spectral contrast \cite{lightner2023understanding}. The relevant point is that the phase-dependent modulation persisted despite these amplitude differences. This persistence motivated the Fourier-phase analysis below, where the global angular phase structure, rather than the absolute yield, is used to identify the topology-carrying component.

Instead of relying on integrated intensities, we can directly access the topology of the emitted field by Fourier transforming the recorded third-harmonic images with respect to the two-colour delay (Fig.~\ref{fig:spectral_control}d). Comparing the angular phase patterns associated with the $m=+1$ Fourier component reveals a topological charge $C=1$ for both enantiomers, in agreement with the symmetry analysis. Although the experimental phase map of the right enantiomer contains three singularities, two carry opposite charge and therefore cancel. These additional vortex points arise from experimental imperfections and do not alter the net topology of the signal. The corresponding TDSE simulations reproduce both the topological charge and the $\pi$-phase offset between opposite enantiomers (Fig.~\ref{fig:spectral_control}e), which is also directly visible in the experimental phase maps: for example, the 12 o'clock direction corresponds to phase $\pi$ for the left enantiomer and phase 0 for the right enantiomer.

The simulations incorporate realistic laser pulses and reproduce the experimentally observed characteristic vortex-centre displacement arising from residual beam misalignment (Fig.~\ref{fig:spectral_control}e). Crucially, while static morphology, collection asymmetries, and local intensity fluctuations can distort the emission envelope, they cannot destroy the phase-locked angular evolution of the signal. The enantiomeric information is therefore robustly encoded in the topology of the angular response rather than in the absolute signal intensity.

Simulations for different molecules indicate that the topologically protected phase maps provide direct access to the molecule-specific phase difference between chiral and achiral pathways. The topology remains unchanged, whereas the entire map undergoes a molecule-dependent rotation, providing a robust fingerprint of molecular identity encoded in its quantum phase (Fig.~\ref{fig:spectral_control}f).

Our results show that latent light topology can become encoded in the chiral non-linear optical response, providing a robust route to chiral metrology in complex and disordered environments. By establishing the first experimental realization of chiral topological light and synthetic chiral light in randomly oriented chiral powders, we extend this concept beyond atomic gases \cite{kohnke2025multiphoton} to condensed matter and demonstrate its viability as a platform for electric-dipole-driven all-optical chiral discrimination. By demonstrating that topology can protect information generated through chiral light--matter interactions, this work introduces a new paradigm in which chirality is encoded, controlled, and detected through topological observables.

\newpage
\section*{Methods}
\textbf{Experimental details:}
\paragraph{Overall optical layout.}
The measurements were performed using a Ti:sapphire femtosecond laser system followed by an optical parametric amplifier, which provided synchronized near-infrared pulses for trefoil-field generation. In the main trefoil configuration, the \(2000~\mathrm{nm}\) output was used as the fundamental \(\omega\) beam and was frequency-doubled in a type-1 BBO crystal with 500 $\mu m$ thickness to generate the \(1000~\mathrm{nm}\) \(2\omega\) beam. The residual fundamental and the second-harmonic beam were separated and recombined with dichroic mirrors to form a collinear two-colour field. A fused-silica wedge pair controlled the relative \(\omega\)-\(2\omega\) phase, while superachromatic wave plates set the two beams to opposite helicities, thereby producing the trefoil electric-field waveform before focusing.

The synthesized two-colour field was directed to the sample and focused with a long-working-distance objective. For chiral perovskite measurements, the samples were mounted in a vacuum stage to suppress sample degradation and environmental fluctuations. The same type of objective collected the emitted nonlinear radiation. Residual driving light was attenuated by optical filters, the emission was sent either to a thermoelectrically cooled CCD camera for Fourier-plane imaging or coupled into a fiber spectrometer for wavelength-resolved detection. 

\paragraph{Materials.}
Lead iodide (\(\mathrm{PbI_2}\), 99.999\%), hydriodic acid (HI, 57 wt.\% in \(\mathrm{H_2O}\)) and hypophosphorous acid (\(\mathrm{H_3PO_2}\), 50 wt.\% in \(\mathrm{H_2O}\)) were purchased from Aladdin. 1-(3-bromophenyl)-ethylamine, \((R/S)\)-3Br-MBA, was purchased from Macklin. All reagents were used as received without further purification unless otherwise stated.

\paragraph{Synthesis of chiral perovskite crystals.}
Chiral perovskite single crystals were synthesized using a modified cooling method. In brief, \(\mathrm{PbI_2}\) powder (\(231~\mathrm{mg}\)) was dissolved in a mixed solvent of HI (\(6~\mathrm{ml}\)) and \(\mathrm{H_3PO_2}\) (\(0.5~\mathrm{ml}\)). The solution was stirred vigorously at \(100~^{\circ}\mathrm{C}\) for \(1~\mathrm{hour}\) to promote complete dissolution. Subsequently, \((R/S)\)-3Br-MBA (\(151~\mu\mathrm{l}\)) and deionized water (\(1~\mathrm{ml}\)) were slowly added dropwise into the solution. The mixture was further stirred at \(100~^{\circ}\mathrm{C}\) for \(4~\mathrm{hour}\). The resulting solution was transferred to an oven with a programmed temperature controller and kept at \(100~^{\circ}\mathrm{C}\) for \(0.5~\mathrm{hour}\). The temperature was then reduced to \(90~^{\circ}\mathrm{C}\) at a rate of \(1~^{\circ}\mathrm{C}~\mathrm{h}^{-1}\), followed by slow cooling to room temperature at a rate of \(0.5~^{\circ}\mathrm{C}~\mathrm{h}^{-1}\). The obtained crystals were removed from the mother solution, rinsed with anhydrous ether to remove residual solution, and dried under vacuum for \(12~\mathrm{hours}\). The molecular structure of \((R/S)\)-3Br-MBA and the image of the synthesized perovskite crystal can be found in Supplementary Information.

\paragraph{Fabrication of chiral perovskite nanoflakes and films.}
Chiral perovskite nanoflakes were prepared by mechanically exfoliating the bulk single crystals with adhesive tape. For film preparation, chiral perovskite single crystals (\(0.2~\mathrm{mmol}\)) were dissolved in dimethylformamide (DMF, \(1~\mathrm{ml}\)) to form a precursor solution. The precursor solution was spin-coated on quartz substrates at \(2000~\mathrm{rpm}\) for \(30~\mathrm{s}\). The resulting films were annealed at \(70~^{\circ}\mathrm{C}\) for \(5~\mathrm{min}\). The whole deposition process was carried out in a nitrogen-filled glove box with \(\mathrm{O_2}<0.5~\mathrm{ppm}\) and \(\mathrm{H_2O}<0.5~\mathrm{ppm}\).  The qualities of the fabricated perovskite nanoflakes and films are characterized by XRD, as shown in Supplementary Information.

\paragraph{Linear optical and structural characterizations.}
Bright-field optical images were acquired with an optical microscope (LEICA DM750M). Steady-state photoluminescence spectra were measured using a XploRa Raman system (Horiba) with a \(473~\mathrm{nm}\) excitation laser. Absorption spectra were recorded with a Hitachi U-3900. The crystal structure was characterized by X-ray diffraction (Smartlab (3), Rigaku) using Cu \(K_{\alpha}\) radiation with a wavelength of \(0.15406~\mathrm{nm}\). Circularly polarized photoluminescence measurements were performed under \(405~\mathrm{nm}\) laser excitation. The emitted photoluminescence was collected by a spectrometer after passing through a quarter-wave plate, a linear polarizer and a long-pass filter. The two circular polarization components were obtained by changing the relative angle between the quarter-wave plate and the polarizer. A weak excitation power of approximately \(100~\mathrm{nW}\) was used to avoid laser-induced degradation or ablation. UV-Absorption and CPL spectroscopy are shown in Supplementary Information.

\paragraph{Laser source and pulse characterization.}
The nonlinear optical experiments were driven by a Ti:sapphire femtosecond laser system  (Spectra Physics Solstice Ace) followed by an optical parametric amplifier (Spectra Physics TOPAS Prime-F). The TOPAS provided synchronized near-infrared pulses at \(1340~\mathrm{nm}\) and \(2000~\mathrm{nm}\). The pulse durations of the \(1340~\mathrm{nm}\) and \(2000~\mathrm{nm}\) outputs were approximately \(77~\mathrm{fs}\) and \(100~\mathrm{fs}\), respectively, as characterized by frequency-resolved optical gating (FROG) (see Supplementary Information). The \(2000~\mathrm{nm}\) output was used as the fundamental branch for trefoil-field synthesis.  The 800 nm pump laser with 34 fs pulse duration was  also characterized by FROG (see Supplementary Information).

\paragraph{Polarization control and beam calibration.}
The polarization states of the two-colour components were controlled with superachromatic half-wave and quarter-wave plates (Thorlabs SAHWP05M-1700 and SAQWP05M-1700). Here SA denotes superachromatic wave plates, which were used to minimize wavelength-dependent polarization errors between the \(2000~\mathrm{nm}\) and \(1000~\mathrm{nm}\) fields. All wave plates were mounted on motorized rotation stages (Thorlabs PRM1), allowing fine adjustment of the ellipticity and relative polarization angle. The ellipticity of each colour component was measured with a motorized polarizer. The focused pump spot was relayed by a telescope imaging system and monitored with a DataRay beam profiler (WinCamD-LCM) to determine the beam size and spatial quality at the sample plane (see Supplementary Information).

\paragraph{Two-colour trefoil-field generation.}
For trefoil-field measurements, the \(2000~\mathrm{nm}\) pulse was frequency-doubled in a 500 $\mu m$ beta-barium borate (BBO) crystal (HG Optronics.,INC.) to generate the \(1000~\mathrm{nm}\) component. The residual \(2000~\mathrm{nm}\) field and the \(1000~\mathrm{nm}\) second-harmonic field were recombined to form a phase-locked \(\omega\)-\(2\omega\) two-colour field. The two-colour components were converted into counter-rotating circularly polarized fields, whose coherent superposition generated a three-lobed trefoil waveform in the transverse plane. 

The relative phase of the two-colours was controlled with a fused-silica wedge pair. The wedges had an apex angle of \(9^{\circ}\), with a minimum thickness of \(300~\mu\mathrm{m}\). For fine phase control, one wedge was mounted on a piezoelectric translation stage  (on Nators LS-1730), enabling sub-wavelength tuning of the relative \(\omega\)-\(2\omega\) phase.  This configuration allowed the two-colour phase to be scanned while maintaining beam pointing, spatial overlap and pulse quality. The two-colour field was naturally phase-locked to ensure reproducible phase-dependent measurements.  The influence of active beam-stability has been compared in Supplementary Information.

\paragraph{Sample environment and excitation geometry.}
The measurements were performed on chiral hybrid perovskite powder thin films. The samples were first spin-coated and protected by PDMS. Then the samples were mounted in a vacuum cooling stage (Customized version of Instec HCP421V-PM) and kept under vacuum but room temperature during the measurements to protect the sample surface and suppress degradation or phase transition under optical excitation. The phase-locked trefoil beam was focused onto the sample with a Mitutoyo long-working-distance achromatic objective (MY20X-824). The same objective was used to collect the emitted nonlinear radiation. In addition, control measurements obtained using an 800 nm + 400 nm trefoil field with an 800 nm probe are shown in Supplementary Information. In these measurements, the spectra were recorded using a fiber-coupled spectrometer without recording images. To minimize chromatic aberration during both focusing and signal collection over the broad spectral range, a reflective objective was used for both excitation and collection (Thorlabs LMM40X-UVV). 

\paragraph{Nonlinear emission detection.}
The optical handedness of the trefoil field was reversed by changing the motorized SAQWP. 
The nonlinear emission was recorded by using both angular imaging and spectral detection channels. For angular imaging, the emitted light was projected onto a thermoelectrically cooled CCD camera (LBtek HS200RM-U). In the CCD detection arm, residual driving fields were suppressed using a band-pass filter centred near \(650~\mathrm{nm}\) (LBtek MBP25-650-50), together with a Thorlabs NENIR50A filter. For spectral measurements, the nonlinear signal was coupled into a thermoelectrically cooled fiber spectrometer (Oceanhood DQPRO). In the spectrometer arm, two Thorlabs NENIR50A filters were used to attenuate residual near-infrared driving light before the spectrometer entrance. For each two-colour phase, the angular emission pattern and the wavelength-resolved nonlinear spectrum were recorded under fixed focusing and collection conditions.

\textbf{Theoretical details:}

\paragraph{Molecular geometry.}Theoretical support for the experiment was carried out using a molecular carvone over chiral perovskite crystal as it requires significantly less computational power and complexity to calculate transition dipole moments. As our proposed method is not molecule dependent, we expect no qualitative difference between the two setups. 

The molecular geometry of the most stable R-carvone conformer was taken from \cite{lambert_optical_2012}. It was reoptimized using Density Functional Theory (DFT) \cite{kohn_density_1996,ziegler_approximate_2002} with the B3LYP \cite{lee_development_1988,becke_density-functional_1988} functional and the 6-311G(d,p) \cite{mclean_contracted_2008} basis set available in Gaussian 16 \cite{g16}. Time-dependent DFT with CAM B3LYP/d-aug-cc-pVDZ \cite{CAMB3LYP,CAMB3LYP_2,augccpvdz_2,augccpvdz_3} was used to evaluate the excitation energies and transition dipoles for the first 100 excited states. The transition dipoles between the excited states were computed with the Multiwfn software \cite{Multiwfn}.

\paragraph{Numerical simulations.} Nonlinear response of randomly oriented chiral medium was numerically calculated using time-dependent Schr\"{o}dinger equation (TDSE) in the basis set of the field-free eigenstates. The external driving field consisted of two counter-rotating, phase-locked, elliptically polarized beams

\begin{equation}
\mathbf E_{\perp}(t)=E_{\omega}\,\mathbf e_{+}e^{-i\omega t}+E_{2\omega}\,\mathbf e_{-}e^{-i(2\omega t+\delta)}+\mathrm{c.c.},
\label{driving_field}
\end{equation}

where 

\begin{equation}
    \mathbf{e_{\pm}} = \frac{\mathbf{\hat{x}} + i\sigma_{\pm}\varepsilon\mathbf{\hat{y}}}{\sqrt{1 + \varepsilon^2}}
\end{equation}

denote opposite elliptical polarizations with $\sigma_{+} = 1$ for right-handed and $\sigma_{-} = -1$ for left-handed light, and $\delta$ is the two-colour phase delay.  For driving fields, we chose wavelengths $\lambda_{+}=800~\mathrm{nm}$ and $\lambda_{-}=400~\mathrm{nm}$, equal intensities $I_{+}=I_{-}=10^{12}~\mathrm{W/cm^2}$, pulse durations $\mathrm{FWHM}_{+}=\mathrm{FWHM}_{-}=12.5~\mathrm{fs}$ (in field amplitude) and ellipticity $\varepsilon = 0.95$ for both beams. 

The elliptically polarized basis vectors can equivalently be written as a coherent superposition of two counter-rotating circularly polarized components. Using $\mathbf{\tilde{e}}_{\pm}=(\mathbf{\hat{x}}+\sigma_{\pm} i\mathbf{\hat{y}})/\sqrt{2}$, we obtain

\begin{equation}
\mathbf{e}_{\pm}
=
\frac{(1\pm\varepsilon)\mathbf{\tilde{e}}_{+} + (1\mp\varepsilon)\mathbf{\tilde{e}}_{-}}
{\sqrt{2(1+\varepsilon^2)}}.
\end{equation}

Thus, finite ellipticity introduces a weak circular component with the opposite handedness, whose relative amplitude is $(1-\varepsilon)/(1+\varepsilon)$. In the circular limit, $\varepsilon=1$, this counter-rotating component vanishes and $\mathbf{e}_{\pm}$ reduces to a purely circularly polarized field.

To quantify the angular chiral dichroism as a function of the two-colour phase delay, we define

\begin{equation}
CD(\theta, \delta)
=
2
\frac{
\int_{\mathcal{M}} d\rho \rho \left[I_R(\rho,\theta,\delta)-I_S(\rho,\theta,\delta)\right]
}{
\int_{\mathcal{M}} d\rho \rho \left[I_R(\rho,\theta,\delta)+I_S(\rho,\theta,\delta)\right]
} ,
\end{equation}

where $\mathcal{M}$ denotes the region in which either enantiomer signal exceeds a fixed fraction of its peak intensity ($5\%$). This masking suppresses artificially large CD values arising from regions where the total signal is weak.

The intensity can be written as a sum of relevant multiphoton pathways

\begin{equation}
    I_{R,S}(\omega) = |\mathbf{P}_a + \eta\mathbf{P}_c|^2
    \label{eq:intensity}
\end{equation}

Here $\eta_R = -\eta_S = 1$ and $\mathbf{P}_a$ and $\mathbf{P}_c$ denote the achiral and chiral channels respectively. For multiphoton diagrams from the main text

\begin{equation}
\mathbf{P}_c(3\omega)
=
\chi^{(2)}(3\omega)
\left[
\mathbf{E}_{\omega, z} \times \mathbf{E}_{2\omega, -}
+
\mathbf{E}_{\omega, +} \times \mathbf{E}_{2\omega, z}
\right]
\end{equation}

and 

\begin{align}
&\mathbf{P}^{circ}_a(3\omega)
=
\chi^{(3)}(3\omega)
\left[
\left(
\mathbf{E}_{2\omega, z}\cdot \mathbf{E}_{2\omega, z}
\right)
\mathbf{E}^{*}_{\omega, +}
+
\left(
\mathbf{E}_{\omega, z}\cdot \mathbf{E}_{\omega, z}
\right)
\mathbf{E}_{\omega, +}
\right]
\\
&\mathbf{P}^{ell}_a(3\omega)
=
\chi^{(3)}(3\omega)
\left[
\left(
\mathbf{E}_{2\omega, -}\cdot \mathbf{\overline{E}}^{*}_{\omega, -}
\right)
\mathbf{E}_{2\omega, -}
+
\left(
\mathbf{E}_{2\omega, -}\cdot \mathbf{\overline{E}}_{2\omega, +}
\right)
\mathbf{E}^{*}_{\omega, +}
+
\left(
\mathbf{E}_{\omega, +}\cdot \mathbf{\overline{E}}_{\omega, -}
\right)
\mathbf{E}_{\omega, +}
\right]
\label{eq:achiral}
\end{align}

Here achiral channels differ depending on whether the elliptical component, denoted by $\overline{\mathbf{E}}$, is present. Since the topological charge is encoded in the phase of these field components, it is preserved only in the interference cross terms in Eq.~\ref{eq:intensity}. Therefore, by taking the difference between the intensities generated by the two enantiomers, this phase information can be isolated:

\begin{equation}
    I_R - I_S \propto \cos(C\theta).
\end{equation}

Here, $C$ is the total topological charge associated with the achiral--chiral interference term. For circularly polarized driving fields, $C=\pm 3$, whereas for elliptically polarized fields, $C=\pm 1$.

\paragraph{Fourier filtering.} A closer look at multiphoton pathways reveals an additional correlation. The terms in Eq.~ \ref{eq:intensity} can be grouped according to the net number of absorbed $2\omega$ photons. For both circularly and elliptically polarized driving fields, all achiral--chiral interference terms have a net absorption of one $2\omega$ photon. By varying the two-colour phase delay $\delta$ and Fourier transforming over it, this group of terms can be uniquely isolated:

\begin{equation}
    \tilde{I}_{R,S}(m) = \int_{0}^{2\pi} d\delta I_{R,S}(\delta) e^{i\delta m}
\end{equation}

where $m$ denotes the Fourier order associated with the net number of $2\omega$ photons. The phase of the $m=1$ Fourier component then reveals the total topological charge $C$.

\paragraph{Focusing formalism.} Within the paraxial limit, the Lissajous curve of the field traces a planar trajectory in time, leaving the topological structure of the nonlinear response hidden in non-propagating components. Under tight focusing, this planar structure can be lifted out of the plane to form a chiral curve in time, enabling a far-field chiral response. To model tight focusing, we implemented the Richards-Wolf formalism \cite{richards1959electromagnetic}, in which the field at the focus is expressed as

\begin{equation}
\mathbf{E}(x,y,z)
=
\frac{i f\, e^{-i k f}}{2\pi}
\iint_{k_{x}^2 + k_{y}^2 \le k^2}
\mathbf{P}(\theta, \phi)\cos(\theta)\mathbf{E}_{\perp}(k_{x}, k_{y})\,
e^{i \left( k_{x} x + k_{y} y + k_{z} z \right)}
\frac{d k_{x}\, d k_{y}}{k_{z}}.
\label{richard-wolf}
\end{equation}

Here, $f$ is the focal length, $\mathbf{E}_{\perp}$ contains both the polarization and amplitude distribution of the plane wavefront incident on the lens, $\cos(\theta)$ is given by the energy conservation, and $\mathbf{P}(\theta,\phi)$ is the rotation matrix that accounts for the rotation of the electric field as it passes through the lens. This matrix is given by

\begin{equation}
\mathbf{P}(\theta,\phi)
=
\begin{pmatrix}
1 + (\cos\theta - 1)\cos^2\phi 
&
(\cos\theta - 1)\sin\phi\cos\phi
\\[6pt]
(\cos\theta - 1)\sin\phi\cos\phi
&
1 + (\cos\theta - 1)\sin^2\phi
\\[6pt]
-\sin\theta\cos\phi
&
-\sin\theta\sin\phi
\end{pmatrix}.
\end{equation}

The numerical aperture of the focusing lens was chosen to give focal waist radii of $w_{+}=2.6~\mu\mathrm{m}$ and $w_{-}=1.4~\mu\mathrm{m}$.

In the first-order post-paraxial approximation, the longitudinal component acquires orbital angular momentum (OAM) proportional to the handedness of the circular polarization ($\sigma_{\pm}$),

\begin{equation}
    E^{\pm}_{z} \propto e^{i\sigma_{\pm}\phi}
\end{equation}

where $\sigma_{\pm} = \pm1$ and $\phi$ is the azimuthal angle in the traverse plane. Further experimental conditions were modeled by introducing a deformation of the spatial Gaussian beam profiles and beam misalignment directly to $E_{\perp}$ in eq. \ref{richard-wolf}. These parameters were matched to the experimental values.

\paragraph{Orientational averaging.} To describe a randomly orientated ensemble, we ran TDSE simulations for 570 molecular orientations and averaged the induced polarization
\begin{equation}
    \boldsymbol{P} = \frac{1}{8\pi^{2}}\int_{0}^{2\pi}\int_{0}^{2\pi}\int_{0}^{\pi}\boldsymbol{P}_{\chi\phi\theta}\sin(\theta)d\chi d\phi d\theta
\end{equation}
where $\chi, \phi, \theta$ are the Euler angles and $\boldsymbol{P}_{\chi\phi\theta}$ is the induced polarization in the laboratory frame for a given molecular orientation.

The integral over $\phi$, $\theta$ was approximated using the Lebedev quadrature \cite{Lebedev1999AQF} of order 9, which gives 38 points on a sphere ($\phi$, $\theta$).
Integration over $\chi$ was performed using the trapezoid method including 15 rotations around each Lebedev axis.

\section*{Acknowledgments}
P.Z. acknowledges the support by the National Natural Science Foundation of China No.12450401 and No.12474261. P.Z. acknowledges the helpful discussions with T.T.Luu. The experimental work was carried out at the Synergetic Extreme Condition User Facility (SECUF, \url{https://cstr.cn/31123.02.SECUF}). O.S. and J.T. acknowledge ERC-2021-AdG
project ULISSES, grant agreement No 101054696. Views
and opinions expressed are however those of the author(s)
only and do not necessarily reflect those of the Euro-
pean Union or the European Research Council. Nei-
ther the European Union nor the granting authority
can be held responsible for them

\section*{Author contributions}
O.S. and P.Z. conceived the study. G.L. conducted the experimental measurements with the support of T.D.T.; G.L. and J.T. performed the data analysis with the support of P.Z., H.T. and Z.W.; J.T. performed numerical simulations and analytical analysis with the support of M.I. and O.S.;  Y.Z. Q.L. G.L. and J.L. contributed to the chiral perovskite materials development and characterization. P. V.-Z. and D.A. provided dipole matrix elements for carvone molecules, S.P. provided dipole matrix elements for binol molecules. Y.P. contributed to the creation of the experimental setup schematic. L.L. maintained the laser system. O.S. supervised the theoretical part of the study, P.Z. supervised the experimental part of the study. G.L., J.T., P.Z. and O.S. wrote the manuscript with contributions from all co-authors.  
\section*{Competing interests}
The authors declare no competing interests.

\bibliographystyle{unsrtnat}
\bibliography{refs}

\end{document}